\documentclass[preprint,aps,prd,showpacs,superscriptaddress,nofootinbib,tightenlines]{revtex4}

\usepackage{amsfonts}
\usepackage{multirow}
\usepackage{mathrsfs}
\usepackage{graphicx}
\usepackage{amsmath}
\usepackage{amssymb}
\usepackage{bm}
\usepackage{bbm}
\usepackage{color}
\usepackage{array}

\def\OMIT#1{}

\newcommand{\nn}{\nonumber}

\newcommand{\beq}{\begin{equation}}
\newcommand{\eeq}{\end{equation}}
\newcommand{\bqa}{\begin{eqnarray}}
\newcommand{\eqa}{\end{eqnarray}}

\begin{document}
%\preprint{}
%%%%%%%%%%%%%%%%%%%%%%%%%%%%%%%%%%%%%%%%%%%%%%%%%%%%%%%%%%%%%%%%%%%%%%%%%%%%%%
\title{\mbox{}\\[10pt]
Next-to-leading-order QCD corrections to heavy quark fragmentation into
${}^1S^{(1,8)}_0$ quarkonia
}
%%%%%%%%%%%%%%%%%%%%%%%%%%%%%%%%%%%%%%%%%%%%%%%%%%%%%%%%%%%%%%%%%%%%%%%%%%%%%%

\author{Feng Feng\footnote{f.feng@outlook.com}}
\affiliation{China University of Mining and Technology, Beijing 100083, China\vspace{0.2cm}}
\affiliation{Institute of High Energy Physics, Chinese Academy of
Sciences, Beijing 100049, China\vspace{0.2cm}}

\author{Yu Jia\footnote{jiay@ihep.ac.cn}}
\affiliation{Institute of High Energy Physics, Chinese Academy of
Sciences, Beijing 100049, China\vspace{0.2cm}}
\affiliation{School of Physics, University of Chinese Academy of Sciences,
Beijing 100049, China\vspace{0.2cm}}

\author{Wen-Long Sang~\footnote{wlsang@ihep.ac.cn}}
 \affiliation{School of Physical Science and Technology, Southwest University, Chongqing 400700, China\vspace{0.2cm}}

\date{\today}
%%%%%%%%%%%%%%%%%%%%%%%%%%%%%%%%%%%%%%%%%%%%%%%%%%%%%%%%%%%%%%%%%%%%%%%%%%%%%%
\begin{abstract}
Within NRQCD factorization framework, in this work we compute,  at the lowest order in velocity expansion,
the next-to-leading-order (NLO) perturbative corrections to the short-distance coefficients associated with
heavy quark fragmentation into the ${}^1S_0^{(1,8)}$ components of a heavy quarkonium.
Starting from the Collins and Soper's operator definition of the quark fragmentation function, we apply
the sector decomposition method to facilitate the numerical manipulation.
It is found that the NLO QCD corrections have a significant impact.
\end{abstract}

\maketitle

The large-$p_\perp$ production of an identified hadron $H$ in hight-energy collision is dominated by the so-called
fragmentation mechanism.
According to the celebrated QCD factorization theorem,
the inclusive production rate of the $H$ with large $p_\perp$ at hadron collider can
be cast into the following factorized form~\cite{Collins:1989gx}:
%----------------------
\beq
%----------------------
d\sigma[A+B\to H(P_\perp)+X]  = \sum_i d{\hat\sigma}[A+B\to i(P_\perp/z)+X] \otimes
D_{i \to H}(z,\mu)+{\mathcal O}(m_H^2/P_\perp^2),
%----------------------
\label{QCD:factorization:theorem}
%----------------------
\eeq
%----------------------
where $A$, $B$ represent two colliding hadrons, $d\hat{\sigma}$ denotes the partonic cross section,
and the sum in (\ref{QCD:factorization:theorem}) is extended over all parton species ($i=q,\bar{q},g$).
$D_{i \to H}(z)$ characterizes the fragmentation function (FF), encoding the probability for the parton $i$ to hadronize into
a multi-hadron state that contains the hadron $H$ carrying the fractional light-cone momentum $z$ with respect to the parent
parton $i$.

Fragmentation functions are nonperturbative yet universal objects, which provide essential information about the hadronization mechanism.
Similar to parton distribution functions, the scale dependence of FFs is governed by the celebrated Dokshitzer-Gribov-Lipatov-Altarelli-Parisi (DGLAP)
equation:
%----------------------
\beq
%----------------------
{d \over d \ln \mu^2} D_{i \to H}(z,\mu) =  \sum_{i} \int_z^1 {dy\over y}
P_{j i} (y, \alpha_s(\mu)) D_{j \to H}\left({z\over y},\mu\right),
%----------------------
\label{DGLAP:evolution}
%----------------------
\eeq
%----------------------
with $P_{ij}(y)$ the corresponding splitting kernel.
The $\mu$ dependence of the fragmentation function conspires to
compensate the $\mu$ dependence of $d\hat{\sigma}$ in (\ref{QCD:factorization:theorem}), so that the
physical cross section no longer depends on this artificial scale.

Fragmentation functions for quark and gluon into light hadrons such as $\pi,\rho$, proton, {\it etc.} are genuinely nonperturbative objects, which can only be
extracted from experiments.
There has emerged intensive interest in recent years toward explaining the {\tt LHC} data for heavy quarkonium such as $J/\psi$, $\eta_c$, $\chi_{cJ}$
production at large $p_\perp$, thus a thorough understanding of quarkonium FFs become important.
Fortunately, because the heavy quark mass $m_Q$ is much greater than the intrinsic QCD scale $\Lambda_{\rm QCD}$,
the FFs for a heavy quarkonium need not be the entirely non-perturbative object. In fact,
owing to the weak QCD coupling at the length scale $\sim 1/m_Q$, together with the nonrelativistic nature of heavy quarkonium,
the factorization approach based on the nonrelativistic QCD (NRQCD)~\cite{Bodwin:1994jh} can be invoked to
further factorize the quarkonium FFs as the sum of products of perturbatively-calculable short-distance coefficients (SDCs)
and long-distance yet universal NRQCD matrix elements~\cite{Braaten:1993mp,Braaten:1993rw}.
To be specific, let us take the charm quark fragmentation into a quarkonium $H$ as a concrete example.
The NRQCD factorization theorem indicates that
%----------------------
\beq
%----------------------
D_{c \to H}(z,\mu)  =   {d_1(z,\mu)\over m^3} \langle 0\vert \mathcal{O}_1^{H}(^1S_0)\vert 0 \rangle +   {d_8(z,\mu)\over m^3}
\langle 0\vert \mathcal{O}_8^{H}(^1S_0)\vert 0 \rangle+\cdots.
%----------------------
\label{FF:NRQCD:factorization}
%----------------------
\eeq
%----------------------
For the purpose of this work, we are only interested in the NRQCD production operators with the quantum number $^1S_0$:
%----------------------
\begin{subequations}
%----------------------
\label{NRQCD:production:operators}
%----------------------
\bqa
%----------------------
\mathcal{O}_1^{H}(^1S_0)  &=&
\chi^\dagger \psi  \sum_{X} |H+X\rangle  \langle H+X|
\psi^\dagger \chi, \\
\mathcal{O}_8^{H}(^1S_0)  &=&
\chi^\dagger T^a \psi  \sum_{X} |H+X\rangle \langle H+X|
\psi^\dagger T^a \chi,
%----------------------
\eqa
%----------------------
\end{subequations}
%----------------------
where $\psi(\chi^\dagger)$ annihilates a heavy (anti-)quark, respectively.
$T^a$ ($a=1,\ldots, N_c^2-1$) represents the generators of $SU(N_c)$ group in the fundamental representation.
$d_1(z,\mu)$ and $d_8(z,\mu)$ in \eqref{FF:NRQCD:factorization} signify
the SDCs affiliated with the respective production channels.

During the past two decades, the SDCs associated with various quarkonium fragmentation functions
have been computed in NRQCD factorization framework.
The heavy quark fragmentation into $S$-wave charmonia was computed at the lowest order both in $\alpha_s$ and velocity long ago~\cite{Braaten:1993mp,Falk:1993rj,Ma:1994zt,Ma:2013yla,Hong:2014tma,Bodwin:2014bia}.
The relativistic corrections to the heavy quark fragmenting into the $S$-wave charmonia were addressed in Refs.~\cite{Martynenko:2005sf,Sang:2009zz,Yang:2019gga}.
The perturbative corrections for heavy quark fragmentation into $S$-wave charmonia were evaluated in Refs.~\cite{Sepahvand:2017gup,Zheng:2019gnb,Zheng:2019dfk}.
The heavy quark fragmenting into the $P$-wave quarkonia was investigated in Refs.~\cite{Chen:1993ii,Yuan:1994hn,Ma:1995vi,Jia:2012qx,Ma:2014eja}.
On the other hand, the fragmentation functions for quarkonium with different flavor ($B_c/B_c^*$)
from charm/bottom quark~\cite{Ji:1986zr,Chang:1992bb,Braaten:1993jn} have also been calculated. Very recently, the authors in \cite{Zheng:2021mqr} computed the fragmentation functions for the spin-singlet quarkonium by a quark,
which has a distinct flavor with the constituent quark in the quarkonium.

The gluon fragmentation into the $S$-wave quarkonium was originally calculated in Refs.~\cite{Braaten:1993rw,Braaten:1995cj,Cho:1994gb}.
The analytic expression for the gluon fragmenting into a vector quarkonium at LO in $\alpha_s$ was presented in Ref.~\cite{Zhang:2017xoj}.
The relativistic corrections to fragmentation functions for the $S$-wave quarkonium were computed in Refs.~\cite{Bodwin:2003wh,Bodwin:2012xc,Gao:2016ihc,Zhang:2017xoj}.
The NLO perturbative corrections for gluon fragmenting into $S$-wave quarkonia were studied in
Refs.~\cite{Beneke:1995yb,Braaten:2000pc,Artoisenet:2014lpa,Artoisenet:2018dbs,Feng:2018ulg,Zhang:2018mlo}.
The gluon-to-$P$-wave quarkonia  fragmentation functions  were evaluated in \cite{Braaten:1994kd,Ma:1995ci,Hao:2009fa,Feng:2017cjk,Zhang:2020atv}.

The NLO perturbative corrections to $g\to {}^1S_0^{(1,8)}$ channels of quarkonium have recently been considered in
\cite{Artoisenet:2018dbs, Feng:2018ulg, Zhang:2018mlo}. This provides valuable information for
a thorough understanding of $\eta_c$, $h_c$ and $J/\psi$ production and polarization at large $p_\perp$.
To facilitate a more realistic phenomenological analysis, one should also consider the analogous NLO correction from quark fragmentation.
It is the aim of this work to compute the NLO pertrubative correction for
the heavy quark to ${}^1S_0^{(1,8)}$ channel of charmonia (bottomonia).

Our starting point is the gauge-invariant definition for the quark fragmentation functions pioneered
by Collins and Soper long ago~\cite{Collins:1981uw}. Note that this operator definition was first used by Ma to compute the quarkonium FFs in NRQCD factorization~\cite{Ma:1994zt}.
According to the operator definition~\cite{Collins:1981uw} (see also \cite{Bodwin:2014bia}),
the desired $c$-to-$H$ fragmentation function in QCD reads
%----------------------
\bqa \label{CS:def:Fragmentation:Function}
%----------------------
& & D_{c \to H}(z,\mu) =
\frac{z^{D-3} }{ 2\pi \times 4 \times N_c }
\int_{-\infty}^{+\infty} \!dx^- \, e^{-i k^+ x^-}
%----------------------
\\
%----------------------
&& \times {\rm tr}\left[ n\!\!\!/
\langle 0 | \Psi(0)
\Phi^\dagger(0,0,{\bf 0}_\perp) \sum_{X} |H(P)+X\rangle \langle H(P)+X|
\Phi(0,x^-,{\bf 0}_\perp) \bar{\Psi}(0,x^-,{\bf 0}_\perp) \vert 0 \rangle \right].
%----------------------
\nn
%----------------------
%----------------------
\eqa
%----------------------
The light-cone coordinate $V^\mu=(V^+, V^-, {\bf V}_\perp)$ has been used, and  $n^\mu =(0,1,{\bf 0}_\perp)$ is a null reference 4-vector.
The variable $z$ denotes the fraction of the $+$-momentum carried by $H$ with respect to the charm quark, $k^+ = P^+/z$ is the $+$-component momentum
injected by the charm quark field operator $\Psi(x)$.
$D=4-2\varepsilon$ signifies the space-time dimensions. $\mu$ is the renormalization scale for this composite operator.
The insertion of the intermediate states implies
that in the asymptotic future, one only needs project out those out-states that contain a
charmonium $H$ carrying definite momentum $P^\mu$ plus any unobserved hadrons,
which are collectively denoted by the symbol $X$.

The eikonal factor $\Phi(0,x^-,{\bf 0}_\perp)$ in (\ref{CS:def:Fragmentation:Function}) is the
path-ordered exponential of the gluon field, whose role is to ensure the gauge invariance of the FF:
%----------------------
\beq
%----------------------
\Phi(0,x^-,{\bf 0}_\perp) = \texttt{P} \exp
\left[ i g_s \int_{x^-}^\infty d y^- n\cdot A(0^+,y^-,{\bf 0}_\perp) \right],
%----------------------
\label{Gauge:Link:Definition}
%----------------------
\eeq
%----------------------
where $\texttt{P}$ implies the path-ordering, $g_s$ is the QCD coupling constant, and
$A^\mu$ denotes the matrix-valued gluon field in $SU(N_c)$ fundamental representation.

We can appeal to standard perturbative matching method to determine the SDCs $d_{1,8}(z,\mu)$ in \eqref{FF:NRQCD:factorization},
by replacing the physical quarkonium $H$ by a fictitious one $c\bar{c}({}^1S_0^{(1,8)})$.
For the QCD part of the amplitude, one needs to project the $c\bar{c}$ pair onto the intended spin/orbital/color states.
It is convenient to employ the covariant projector to expedite the calculation~\cite{Petrelli:1997ge}:
%----------------------
\bqa
\label{spin:color:projectors}
%----------------------
\Pi_{1,8} &=& \frac{1}{\sqrt{8m^3}}\left({P\!\!\!\slash\over 2}- m\right)\gamma_5 \left({P\!\!\!\slash\over 2}+m\right) \otimes \mathcal{C}_{1,8},
%----------------------
\eqa
%----------------------
with $P^\mu$ designating the total momentum of the $c\bar{c}$ pair,
and the color projectors $\mathcal{C}_{1,8}$  given by
%----------------------
\begin{subequations}
%----------------------
\bqa
%----------------------
\mathcal{C}_1 &=& { \texttt{1}_c  \over \sqrt{N_c}},
%----------------------
\\
%----------------------
\mathcal{C}_8^a &=& \sqrt{2} T^a.
%----------------------
\eqa
%----------------------
\end{subequations}
%----------------------
$\texttt{1}_c$ is the $N_c$-dimensional unit matrix. Because we are only interested in the LO accuracy in velocity expansion,
we have neglected the relative momentum between $c$ and $\bar{c}$ in \eqref{spin:color:projectors}, consequently we approximate
$P^2$ by $4m^2$.

%----------------------
\begin{figure}[tb]
%----------------------
\centering
%----------------------
\includegraphics[width=0.9\textwidth]{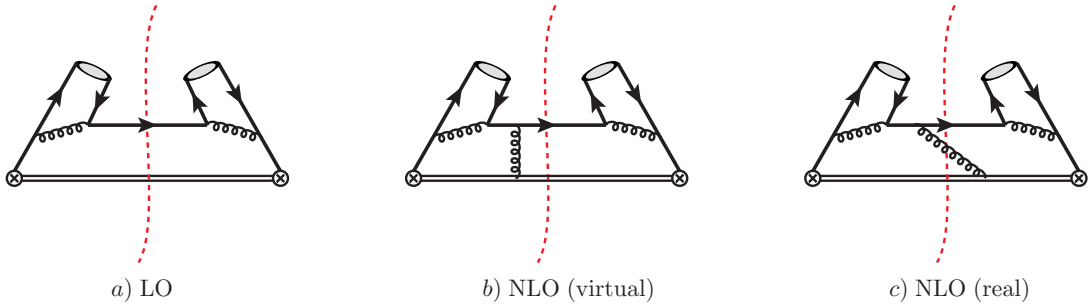}
%----------------------
\caption{Some representative Feynman diagrams for the charm quark fragmentation function $D_{c\to c\bar{c}(^1S_0^{(1,8)})}(z)$
at LO and NLO in $\alpha_s$.
The cap represents the injected quark field operator, while the double line signifies the eikonal line.
%----------------------
\label{Feynman:diagram}}
%----------------------
\end{figure}
%----------------------

Although (\ref{CS:def:Fragmentation:Function}) is manifestly gauge-invariant,  it is simplest to
specialize to the Feynman gauge in practical calculation.
Dimensional regularization is used throughout this work to regularize both UV and IR divergences.
We use two private packages to automatically generate the Feynman diagrams and the associated
cut amplitudes that correspond to the perturbative fragmentation function defined in \eqref{CS:def:Fragmentation:Function}.
One is based on the package {\tt Qgraf}~\cite{Nogueira:1991ex}, and the other based  on {\tt FeynArts}~\cite{Hahn:2000kx}.
We have implemented Feynman rules involving the eikonal propagator and vertex~\cite{Collins:1981uw} as well as those for conventional
QCD propagators and vertices.
Some representative Feynman diagrams for perturbative FF through NLO in $\alpha_s$ are shown in Fig.~\ref{Feynman:diagram}.

With the aid of the covariant projector \eqref{spin:color:projectors}, after obtaining the expressions for the
cut amplitudes, we then utilize the packages {\tt FeynCalc/FormLink}~\cite{Mertig:1990an,Feng:2012tk}
to conduct the Dirac/color trace algebra. We also use the package {\tt Apart}~\cite{Feng:2012iq} to simplify the
amplitude by the method of partial fraction, to make the integrand in loop integrals simpler.

A peculiar structure of the fragmentation function originates from its cut topology,
with insertion of the asymptotic out-state in (\ref{CS:def:Fragmentation:Function}).
The corresponding phase-space integration measure through the cut becomes~\cite{Bodwin:2003wh,Bodwin:2012xc}
%----------------------
\bqa
%----------------------
d\Phi_n &=& {8\pi m \over S_n} \delta(k^+-P^+-\sum_{i=1}^n k_i^+) \prod_{i=1}^n \frac{dk^+_i}{2k_i^+}\frac{d^{D-2}k_{i\perp}}{(2\pi)^{D-1}} \theta(k^+_i),\label{phase:space}
%----------------------
\eqa
%----------------------
where $k_i$ ($i=1,2$) stands for the momentum of the $i$-th on-shell parton that passes through the cut,
and $S_n$ is the statistical factor for $n$ identical partons.
For our purpose, suffices it to set $S_n=1$. It is important to note that integration over $k_i^+$ can be transformed into
a parametric integration in a finite interval, but the integration over the transverse momentum
$k_{i,\perp}$ are utterly unbounded, {\it i.e.}, from $-\infty$ to $+\infty$.
This feature indicates that the integration over $k_{i,\perp}$ could be regarded as
loop integration in $D-2$-dimensional spacetime.

As first noticed in \cite{Feng:2018ulg,Zhang:2018mlo}, some technical nuisance may arise in utilizing the standard integration-by-part (IBP)
technique to tackle the NLO real corrections, whereas it is quite safe to apply IBP to handle NLO virtual correction.
As expounded in \cite{Feng:2018ulg}, in this work we decide to utilize the {\it sector decomposition} technique~\cite{Binoth:2000ps,Binoth:2003ak}
to evaluate all the NLO real correction diagrams,

Through perturbative matching procedure, we able able to deduce the SDCs appearing in
\eqref{FF:NRQCD:factorization} order by order in $\alpha_s$:
%----------------------
\bqa
%----------------------
d_{1,8}(z,\mu) = d_{1,8}^{\rm LO}(z,\mu)  + \frac{\alpha_s(\mu)}{\pi} d_{1,8}^{\rm NLO}(z,\mu) + \cdots.
%----------------------
\eqa
%----------------------

The desired LO SDCs  in $D=4-2\epsilon$ dimensions turn out to be
%----------------------
\begin{subequations}
\label{sdc-LO}
%----------------------
\bqa
%----------------------
d_1^{\rm LO}(z,\mu) &=&\frac{16\alpha_s^2 (1-z)^2 z \, (4\pi)^{\epsilon} (2-z)^{-2\epsilon} \Gamma(1+\epsilon) }{243 (2-z)^6}
\times\Big[ (48+8z^2-8z^3+3z^4) \nonumber\\
&&
-96(1-z)\epsilon + (48-96z+40z^2+8z^3-3z^4)\epsilon^2
\Big]\\
%----------------------
d_8^{\rm LO}(z,\mu) &=&\frac{\alpha_s^2 (1-z)^2 z \, (4\pi)^{\epsilon} (2-z)^{-2\epsilon} \Gamma(1+\epsilon) }{162 (2-z)^6}
\times\Big[ (48+8z^2-8z^3+3z^4) \nonumber\\
&&
-96(1-z)\epsilon + (48-96z+40z^2+8z^3-3z^4)\epsilon^2\Big].
%----------------------
\eqa
%----------------------
\end{subequations}
%----------------------
Setting $\epsilon=0$, our result for $d_1^{\rm LO}$ is compatible with \cite{Braaten:1993mp},
and our result for $d_8^{\rm LO}$ agrees with \cite{Yuan:1994hn}.

Summing both real and virtual NLO corrections, implementing standard renormalization procedure for QCD lagrangian
(the QCD coupling constant is renormalized under the $\overline{\rm MS}$ scheme), we find that the infrared pole indeed
disappear in the NLO SDCs in both color-singlet and octet channels.
However, there still scurvies an  $z$-dependent single UV pole.
This is simply the symptom that the fragmentation function at NLO still requires an additional
operator renormalization~\cite{Collins:1981uw,Bodwin:2014bia}:
%----------------------
\beq
%----------------------
D^{\overline{\rm MS}}_{c\to H}(z,\mu) = D_{g\to H}(z,\mu) -
{1\over \epsilon}{\alpha_s\over 2\pi} \int_z^1 \!\!{dy\over y}\, P_{cc}(y) D_{c\to H}(z/y,\mu)
- {1\over \epsilon}{\alpha_s\over 2\pi} \int_z^1 \!\!{dy\over y}\, P_{gc}(y) D_{g\to H}(z/y,\mu),
%----------------------
\label{DGLAP:renormalization}
%----------------------
\eeq
%----------------------
where $P_{cc}(y)$ and $P_{gc}(y)$ represent the Altarelli-Parisi splitting kernels for $c\to c$ and $c\to g$ respectively:
%----------------------
\beq
%----------------------
P_{cc}(y) = C_F \left[ \frac{1+y^2}{(1-y)_+} + \frac{3}{2}\delta(1-y)\right], \qquad P_{gc}(y) = C_F \frac{1+(1-y)^2}{y}.
%----------------------
\eeq
%----------------------
Note in \eqref{DGLAP:renormalization} the UV pole is subtracted in accordance with the $\overline{\rm MS}$ scheme.

Finally, the renormalized SDCs at NLO in $\alpha_s$ can be parameterized in the following form:
%----------------------
\bqa
%----------------------
d_{1,8}^{\rm NLO}(z,\mu) &=& \beta_0 \ln\frac{\mu_R^2}{m^2} d^{\rm LO}_{1,8}(z)+\frac{1}{2}\ln\frac{\mu^2}{m^2}\bigg[\int_z^1\frac{dy}{y}P_{cc}(y) d^{\rm LO}_{1,8}(z/y)
%----------------------
\nn
%----------------------
\\
%----------------------
&+& \int_z^1\frac{dy}{y}P_{gc}(y) d^{\rm LO}_{g \to H}(z/y)\bigg]+f_{1,8}(z),
%----------------------
\label{renormalized:SDCs}
\eqa
%----------------------
with $\beta_0$ the one-loop QCD $\beta$ function.

With the aid of $D$-dimensional expressions for the LO SDCs in \eqref{sdc-LO}, we can deduce the
coefficients of the $\ln\mu^2$ terms analytically. For the color-singlet channel, we have
%----------------------
\begin{subequations}
%----------------------
\bqa
%----------------------
&&\int_z^1\frac{dy}{y}P_{cc}(y) d^{\rm LO}_{1}(z/y)=\frac{16\alpha_s^2}{10935 (2-z)^6}\bigg[
120 z (48+8 z^2-8 z^3+3 z^4) (1-z)^2 \ln (1-z)\nn\\
&& -15 (5632-16320 z+19008 z^2-10928 z^3+3072 z^4-372 z^5+20 z^6-9 z^7 ) \ln (2-z) \nn\\
&&-15 z ( 192-192 z-272 z^2+288 z^3+36 z^4-76 z^5+21 z^6 ) \ln z \nn\\
&&+2 ( 30848-78352 z+76400 z^2-37160 z^3+10420 z^4-1553 z^5+192 z^6  ) (1-z) \bigg],\\
&&\int_z^1\frac{dy}{y}P_{gc}(y) d^{\rm LO}_{g \to ^1S_0^{[1]}}(z/y)= \frac{\alpha_s^2}{81z}\bigg[
-6 (z+2) z {\rm Li}_2(z) -3 z^2 \ln z +6 (1-z)(2 z+1)\ln(1-z)  \nn\\
&& 2 + (2 \pi ^2-18)z + (\pi^2+12) z^2 +4 z^3
 \bigg].
%----------------------
\eqa
%----------------------
\end{subequations}
%----------------------

For the color-octet channel, the coefficients of $\ln \mu^2$ term read
%----------------------
\begin{subequations}
%----------------------
\bqa
%----------------------
&&\int_z^1\frac{dy}{y}P_{cc}(y) d^{\rm LO}_{8}(z/y)=\frac{\alpha_s^2}{7290 (2-z)^6}\bigg[120 z
(48+8 z^2-8 z^3+3 z^4) (1-z)^2 \ln (1-z)\nn\\
%----------------------
&& -15 ( 5632-16320 z+19008 z^2 -10928 z^3+3072 z^4 -372 z^5 +20 z^6-9 z^7 ) \ln (2-z) \nn\\
&&-15 z ( 192-192 z-272 z^2+288 z^3+36 z^4-76 z^5+21 z^6) \ln z + \nn\\
%----------------------
&& 61696-218400z+309504 z^2-227120 z^3+95160 z^4-23946 z^5 +3490 z^6-384 z^7 \bigg],\\
%----------------------
 &&\int_z^1\frac{dy}{y}P_{gc}(y) d^{\rm LO}_{g \to ^1S_0^{[8]}}(z/y)= \frac{10\alpha_s^2}{432z}\bigg[
-6 (z+2) z {\rm Li}_2(z) -3 z^2 \ln z +6 (-2 z^2+z+1) \ln (1-z) \nn\\
&& +2 +2 (\pi ^2-9) z +(12+\pi ^2) z^2 +4 z^3
 \bigg].
%----------------------
\eqa
%----------------------
\end{subequations}
%----------------------
We notice the occurrence of $1/z$ singularity in the integrals involving $P_{gc}$.

\begin{table}[t]
\caption{\label{c1-table}
Numerical values of non-logarithmic color-singlet coefficient functions $c_{1,2,3}^{(1)}(z)$
as introduced in \eqref{1s0-coeff}.
We caution that the actual values of $c^{(1)}_i(z)$ should be multiplied by an extra factor $10^{-2}$.
}
\newcolumntype{L}{>{$}l<{$}}
\newcolumntype{R}{>{$}r<{$}}
\newcolumntype{C}{>{$}c<{$}}
\begin{ruledtabular}
\begin{tabular}{CRRRCRRR}
 z & c^{(1)}_1(z) & c^{(1)}_2(z) & c^{(1)}_3(z) & z & c^{(1)}_1(z) & c^{(1)}_2(z) & c^{(1)}_3(z)  \\
\hline
 0.05 & -84.7925(6) & 0.26142 & 0.25443 & 0.55 & -18.8763(3) & 0.91199 & -0.45891 \\
 0.10 & -54.2569(8) & 0.42309 &  0.39228& 0.60 & -13.5185(3) & 0.89995 & -0.65036 \\
 0.15 & -45.6338(2) & 0.54973 & 0.47410 & 0.65 & -7.8484(3) & 0.88283 & -0.77734 \\
 0.20 & -41.7388(6) & 0.65275 & 0.50740 & 0.70 & -2.1861(4) & 0.86303 & -0.79912 \\
 0.25 & -39.2176(5) & 0.73655 & 0.49315 & 0.75 & 3.0117(3) & 0.83987 & -0.68278 \\
 0.30 & -36.9564(2) & 0.80310 & 0.43092 & 0.80 & 7.1367(2) & 0.80516 & -0.42365 \\
 0.35 & -34.4680(2) & 0.85342 & 0.32112 & 0.85 & 9.4336(3) & 0.73635 & -0.76013 \\
 0.40 & -31.5042(2) & 0.88828 & 0.16675 & 0.90 & 9.0611(2) & 0.58954 &  2.16314 \\
 0.45 & -27.9368(2) & 0.90861 &-0.02492 & 0.95 & 5.3646(3) & 0.31034 &  0.23972\\
 0.50 & -23.7186(3) & 0.91581 &-0.24051 & 0.99 & 0.6854(2) & 0.03175 & 0.03101 \\
\end{tabular}
\end{ruledtabular}
\end{table}

\begin{table}[t]
\caption{\label{c8-table}
Numerical values of non-logarithmic color-octet coefficient functions $c_{1,2,3}^{(1)}(z)$
as introduced in \eqref{1s0-coeff}.
We caution that the actual values of $c^{(1)}_1(z)$, $c^{(1)}_2(z)$ and $c^{(1)}_3(z)$ should be multiplied by a factor $10^{-2}$.
}
\newcolumntype{L}{>{$}l<{$}}
\newcolumntype{R}{>{$}r<{$}}
\newcolumntype{C}{>{$}c<{$}}
\begin{ruledtabular}
\begin{tabular}{CRRRCRRR}
 z & c^{(1)}_1(z) & c^{(1)}_2(z) & c^{(1)}_3(z) & z & c^{(1)}_1(z) & c^{(1)}_2(z) & c^{(1)}_3(z)  \\
\hline
 0.05 & -89.6874(4) & 0.02451 & 0.02385 & 0.55 & 10.0661(5) & 0.08550 & -0.04302 \\
 0.10 & -26.5678(3) & 0.03967 & 0.03678 & 0.60 & 9.8386(4) & 0.08437 & -0.06097 \\
 0.15 & -7.4488(4) & 0.05154 & 0.04445 & 0.65 & 9.4633(7) & 0.08276 & -0.07288 \\
 0.20 & 0.9549(3) & 0.06120 & 0.04757 & 0.70 & 7.8382(6) & 0.08091 & -0.07492 \\
 0.25 & 5.2463(2) & 0.06905 & 0.04623 & 0.75 & 5.3851(4) & 0.07874 & -0.06401 \\
 0.30 & 7.5974(3) & 0.07529 & 0.04040 & 0.80 & 2.7520(4) & 0.07548 & -0.03972 \\
 0.35 & 8.9190(4) & 0.08001 & 0.03011 & 0.85 & 0.2366(5) & 0.06903 & -0.00713 \\
 0.40 & 9.6491(3) & 0.08328 & 0.01563 & 0.90 & -1.5717(4) & 0.05527 & 0.02028 \\
 0.45 & 10.0142(5) & 0.08518 &-0.00234 & 0.95 & -1.7326(5) & 0.02909 & 0.02247 \\
 0.50 & 10.1334(3) & 0.08586 &-0.02255 & 0.99 & -0.2953(0) & 0.00298 & 0.00291 \\
\end{tabular}
\end{ruledtabular}
\end{table}

The non-logarithmic terms can be decomposed in terms of distinct flavor structure:
%----------------------
\bqa
\label{1s0-coeff}
%----------------------
f_{1,8}(z)=\alpha_s^2\bigg[c^{(1,8)}_1(z)+n_L c^{{1,8}}_2(z)+n_H c^{(1,8)}_3(z)\bigg]
%----------------------
\eqa
%----------------------
where $n_L$ denotes the number of light quarks, and $n_H=1$ is the number of heavy quark.
The numerical values of individual coefficient functions $c^{(1,8)}_i(z)$ ($i=1,2,3$) have been tabulated in
Tables~\ref{c1-table} and \ref{c8-table}.

For a concrete investigation of the heavy quark fragmentation functions, we take the following values for the
one-loop pole mass of charm and bottom quarks
%----------------------
\beq
%----------------------
m_c = 1.4\,\text{GeV},\; m_b=4.6\,\text{GeV}.
%----------------------
\eeq
%----------------------
The running QCD coupling is computed with two-loop accuracy with the aid of the package {\tt RunDec}~\cite{Chetyrkin:2000yt}.
We have taken $n_L=3,4$ for charmonium and bottomonium, respectively so that $n_f=n_L+1$.

%----------------------
\begin{figure}[tb]
%----------------------
\centering
%----------------------
\includegraphics[width=0.45\textwidth]{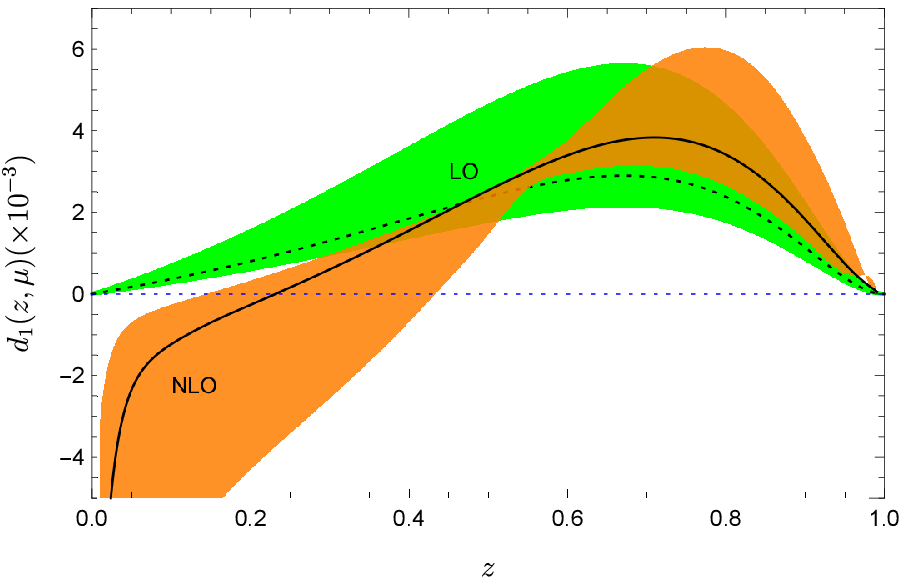} \;
\includegraphics[width=0.45\textwidth]{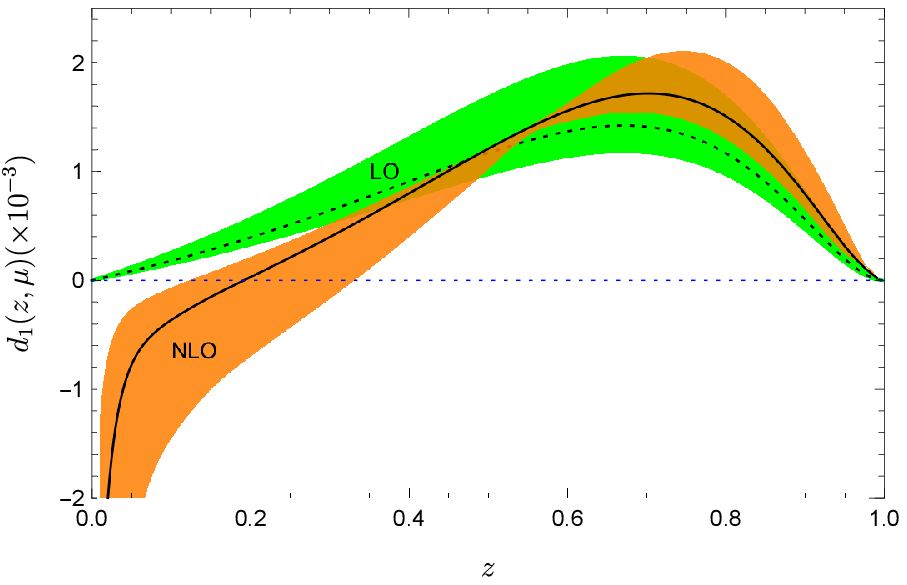}
\includegraphics[width=0.45\textwidth]{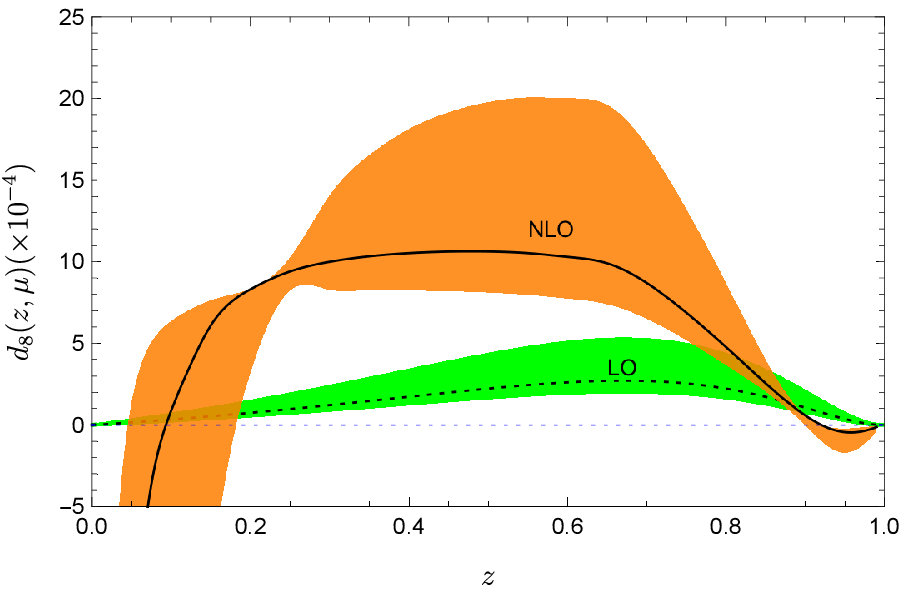} \;
\includegraphics[width=0.45\textwidth]{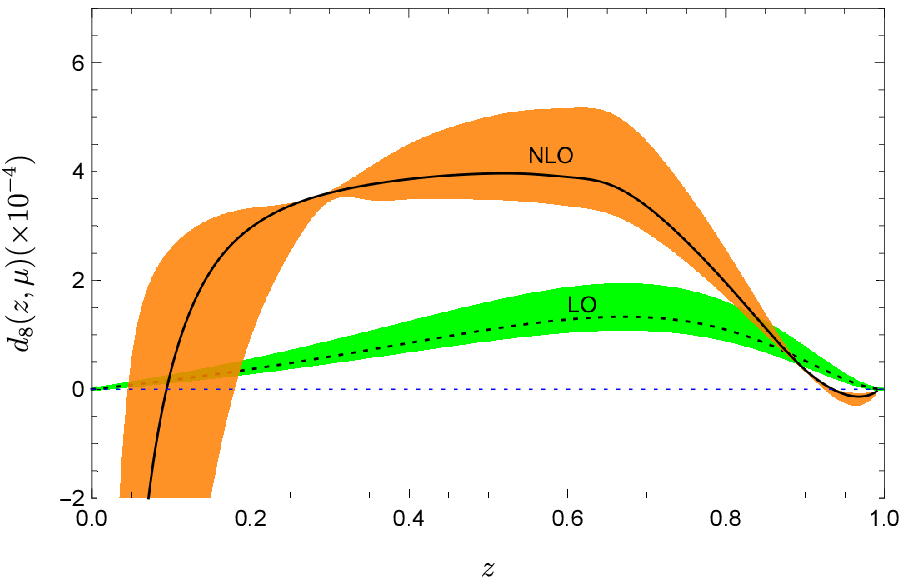}
%----------------------
\caption{The SDCs $d_{1,8}(z)$ associated with heavy quark fragmentation functions, accurate up to NLO in $\alpha_s$.
The two figures in the left column correspond to $d_{1,8}(z)$ for charm quark fragmentation into charmonium, while the two figures in
the right column correspond to those for bottom quark fragmentation.
The uncertainty bands are derived by varying the factorization scale $\mu$ from $m_Q$ to $3m_Q$,
with the central value taken at $2m_Q$.
%----------------------
\label{figDg}}
%----------------------
\end{figure}
%----------------------

The profiles of the SDCs $d_{1,8}(z)$ through the NLO in $\alpha_s$  are displayed in Fig.~\ref{figDg}
for $c$ to charmonium and $b$ to bottomonium, respectively.
We have equated the renormalization scale $\mu_R$ and factorization scale $\mu$
in \eqref{renormalized:SDCs} for simplicity. The central curves are obtained by setting $\mu=2m_Q$, while the
uncertainty bands are obtained by varying $\mu$ from $m_Q$ to $3m_Q$.
Aa can be clearly visualized, the NLO perturbative corrections have a significant impact on both color-singlet and octet
channels, for both charm and bottom fragmentation function, including which
would qualitatively modify the shapes of LO fragmentation functions.
Therefore, it appears mandatory to include the NLO QCD corrections in the future phenomenological analysis.

\noindent{\it Note added.} After this work is completed and while we were preparing the manuscript,
very recently a preprint~\cite{Zheng:2021ylc} has appeared, which also computes the NLO perturbative
corrections to the heavy quark fragmentation into a ${}^1S_0^{(1)}$ quarkonium.
Their numerical results appear to be compatible with ours in this color-singlet channel.

%-----------------------
\begin{acknowledgments}
%{\noindent\it Acknowledgment.}
%-----------------------
 The work of F. F. is supported by the National Natural
Science Foundation of China under Grant No. 11875318, No. 11505285, and by the Yue
Qi Young Scholar Project in CUMTB.
%-----------------------
The work of Y. J. is supported in part by the National Natural
Science Foundation of China under Grants No. 11925506, 11875263, No. 12070131001
(CRC110 by DFG and NSFC).
%-----------------------
The work of W.-L. S. is supported by the National
Natural Science Foundation of China under Grants No. 11975187 and the Natural Science
Foundation of ChongQing under Grant No. cstc2019jcyj-msxmX0479.
%-----------------------
\end{acknowledgments}
%-----------------------


\begin{thebibliography}{99}

%\cite{Collins:1989gx}
\bibitem{Collins:1989gx}
J.~C.~Collins, D.~E.~Soper and G.~F.~Sterman,
%``Factorization of Hard Processes in QCD,''
Adv. Ser. Direct. High Energy Phys. \textbf{5}, 1-91 (1989)
doi:10.1142/9789814503266\_0001
[arXiv:hep-ph/0409313 [hep-ph]].
%1288 citations counted in INSPIRE as of 03 Jun 2021

%\cite{Bodwin:1994jh}
\bibitem{Bodwin:1994jh}
G.~T.~Bodwin, E.~Braaten and G.~P.~Lepage,
%``Rigorous QCD analysis of inclusive annihilation and production of heavy quarkonium,''
Phys. Rev. D \textbf{51}, 1125-1171 (1995)
[erratum: Phys. Rev. D \textbf{55}, 5853 (1997)]
doi:10.1103/PhysRevD.55.5853
[arXiv:hep-ph/9407339 [hep-ph]].
%2537 citations counted in INSPIRE as of 03 Jun 2021

%\cite{Braaten:1993mp}
\bibitem{Braaten:1993mp}
E.~Braaten, K.~m.~Cheung and T.~C.~Yuan,
%``Z0 decay into charmonium via charm quark fragmentation,''
Phys. Rev. D \textbf{48}, 4230-4235 (1993)
doi:10.1103/PhysRevD.48.4230
[arXiv:hep-ph/9302307 [hep-ph]].
%272 citations counted in INSPIRE as of 03 Jun 2021


%\cite{Falk:1993rj}
\bibitem{Falk:1993rj}
A.~F.~Falk, M.~E.~Luke, M.~J.~Savage and M.~B.~Wise,
%``Heavy quark fragmentation to polarized quarkonium,''
Phys. Lett. B \textbf{312}, 486-490 (1993)
doi:10.1016/0370-2693(93)90986-R
[arXiv:hep-ph/9305260 [hep-ph]].
%36 citations counted in INSPIRE as of 03 Jun 2021

%\cite{Ma:1994zt}
\bibitem{Ma:1994zt}
J.~P.~Ma,
%``Calculating fragmentation functions from definitions,''
Phys. Lett. B \textbf{332}, 398-404 (1994)
doi:10.1016/0370-2693(94)91271-8
[arXiv:hep-ph/9401249 [hep-ph]].
%45 citations counted in INSPIRE as of 03 Jun 2021

%\cite{Ma:2013yla}
\bibitem{Ma:2013yla}
Y.~Q.~Ma, J.~W.~Qiu and H.~Zhang,
%``Heavy quarkonium fragmentation functions from a heavy quark pair. I. $S$ wave,''
Phys. Rev. D \textbf{89}, no.9, 094029 (2014)
doi:10.1103/PhysRevD.89.094029
[arXiv:1311.7078 [hep-ph]].
%58 citations counted in INSPIRE as of 03 Jun 2021

%\cite{Hong:2014tma}
\bibitem{Hong:2014tma}
H.~Zhang,
``QCD factorization for heavy quarkonium production and fragmentation functions,''
%0 citations counted in INSPIRE as of 03 Jun 2021

%\cite{Bodwin:2014bia}
\bibitem{Bodwin:2014bia}
G.~T.~Bodwin, H.~S.~Chung, U.~R.~Kim and J.~Lee,
%``Quark fragmentation into spin-triplet $S$-wave quarkonium,''
Phys. Rev. D \textbf{91}, no.7, 074013 (2015)
doi:10.1103/PhysRevD.91.074013
[arXiv:1412.7106 [hep-ph]].
%16 citations counted in INSPIRE as of 03 Jun 2021

%\cite{Ji:1986zr}
\bibitem{Ji:1986zr}
C.~R.~Ji and F.~Amiri,
%``Perturbative {QCD} Predictions for Inclusive Production of Heavy Mesons in $e^+ e^-$ Annihilation,''
Phys. Rev. D \textbf{35}, 3318 (1987)
doi:10.1103/PhysRevD.35.3318
%48 citations counted in INSPIRE as of 03 Jun 2021

%\cite{Chang:1992bb}
\bibitem{Chang:1992bb}
C.~H.~Chang and Y.~Q.~Chen,
%``The Production of B(c) or anti-B(c) meson associated with two heavy quark jets in Z0 boson decay,''
Phys. Rev. D \textbf{46}, 3845 (1992)
[erratum: Phys. Rev. D \textbf{50}, 6013 (1994)]
doi:10.1103/PhysRevD.46.3845
%173 citations counted in INSPIRE as of 03 Jun 2021

%\cite{Braaten:1993jn}
\bibitem{Braaten:1993jn}
E.~Braaten, K.~m.~Cheung and T.~C.~Yuan,
%``Perturbative QCD fragmentation functions for $B_c$ and $B_{c}$ * production,''
Phys. Rev. D \textbf{48}, no.11, R5049 (1993)
doi:10.1103/PhysRevD.48.R5049
[arXiv:hep-ph/9305206 [hep-ph]].
%212 citations counted in INSPIRE as of 03 Jun 2021

%\cite{Martynenko:2005sf}
\bibitem{Martynenko:2005sf}
A.~P.~Martynenko,
%``Relativistic effects in the processes of heavy quark fragmentation,''
Phys. Rev. D \textbf{72}, 074022 (2005)
doi:10.1103/PhysRevD.72.074022
[arXiv:hep-ph/0506324 [hep-ph]].
%20 citations counted in INSPIRE as of 03 Jun 2021

%\cite{Sang:2009zz}
\bibitem{Sang:2009zz}
W.~l.~Sang, L.~f.~Yang and Y.~q.~Chen,
%``Relativistic corrections to heavy quark fragmentation to S-wave heavy mesons,''
Phys. Rev. D \textbf{80}, 014013 (2009)
doi:10.1103/PhysRevD.80.014013
%13 citations counted in INSPIRE as of 03 Jun 2021

%\cite{Yang:2019gga}
\bibitem{Yang:2019gga}
D.~Yang and W.~Zhang,
%``Relativistic corrections of the fragmentation functions for a heavy quark to $B_c$ and $B_c^{*}$,''
Chin. Phys. C \textbf{43}, no.8, 083101 (2019)
doi:10.1088/1674-1137/43/8/083101
[arXiv:1905.02923 [hep-ph]].
%2 citations counted in INSPIRE as of 03 Jun 2021

%\cite{Sepahvand:2017gup}
\bibitem{Sepahvand:2017gup}
R.~Sepahvand and S.~Dadfar,
%``NLO corrections to $c$- and $b$-quark fragmentation into $j/\psi$ and $\gamma$,''
Phys. Rev. D \textbf{95}, no.3, 034012 (2017)
doi:10.1103/PhysRevD.95.034012
%12 citations counted in INSPIRE as of 03 Jun 2021

%\cite{Zheng:2019gnb}
\bibitem{Zheng:2019gnb}
X.~C.~Zheng, C.~H.~Chang, T.~F.~Feng and X.~G.~Wu,
%``QCD NLO fragmentation functions for c or b\textasciimacron{} quark to Bc or Bc* meson and their application,''
Phys. Rev. D \textbf{100}, no.3, 034004 (2019)
doi:10.1103/PhysRevD.100.034004
[arXiv:1901.03477 [hep-ph]].
%10 citations counted in INSPIRE as of 03 Jun 2021

%\cite{Zheng:2019dfk}
\bibitem{Zheng:2019dfk}
X.~C.~Zheng, C.~H.~Chang and X.~G.~Wu,
%``NLO fragmentation functions of heavy quarks into heavy quarkonia,''
Phys. Rev. D \textbf{100}, no.1, 014005 (2019)
doi:10.1103/PhysRevD.100.014005
[arXiv:1905.09171 [hep-ph]].
%5 citations counted in INSPIRE as of 03 Jun 2021

%\cite{Zheng:2021mqr}
\bibitem{Zheng:2021mqr}
X.~C.~Zheng, Z.~Y.~Zhang and X.~G.~Wu,
%``Fragmentation functions for a quark into a spin-singlet quarkonium: Different flavor case,''
Phys. Rev. D \textbf{103}, no.7, 074004 (2021)
doi:10.1103/PhysRevD.103.074004
[arXiv:2101.01527 [hep-ph]].
%1 citations counted in INSPIRE as of 03 Jun 2021

%\cite{Chen:1993ii}
\bibitem{Chen:1993ii}
Y.~Q.~Chen,
%``Perturbative QCD predictions for the fragmentation functions of the P wave mesons with two heavy quarks,''
Phys. Rev. D \textbf{48}, 5181-5189 (1993)
doi:10.1103/PhysRevD.48.5181
%73 citations counted in INSPIRE as of 03 Jun 2021

%\cite{Yuan:1994hn}
\bibitem{Yuan:1994hn}
T.~C.~Yuan,
%``Perturbative QCD fragmentation functions for production of P wave mesons with charm and beauty,''
Phys. Rev. D \textbf{50}, 5664-5675 (1994)
doi:10.1103/PhysRevD.50.5664
[arXiv:hep-ph/9405348 [hep-ph]].
%78 citations counted in INSPIRE as of 03 Jun 2021

%\cite{Ma:1995vi}
\bibitem{Ma:1995vi}
J.~P.~Ma,
%``Quark fragmentation into p wave triplet quarkonium,''
Phys. Rev. D \textbf{53}, 1185-1190 (1996)
doi:10.1103/PhysRevD.53.1185
[arXiv:hep-ph/9504263 [hep-ph]].
%31 citations counted in INSPIRE as of 03 Jun 2021

%\cite{Jia:2012qx}
\bibitem{Jia:2012qx}
Y.~Jia, W.~L.~Sang and J.~Xu,
%``Inclusive $h_c$ Production at $B$ Factories,''
Phys. Rev. D \textbf{86}, 074023 (2012)
doi:10.1103/PhysRevD.86.074023
[arXiv:1206.5785 [hep-ph]].
%14 citations counted in INSPIRE as of 03 Jun 2021

%\cite{Ma:2014eja}
\bibitem{Ma:2014eja}
Y.~Q.~Ma, J.~W.~Qiu and H.~Zhang,
%``Heavy quarkonium fragmentation functions from a heavy quark pair. II. $P$ wave,''
Phys. Rev. D \textbf{89}, no.9, 094030 (2014)
doi:10.1103/PhysRevD.89.094030
[arXiv:1401.0524 [hep-ph]].
%24 citations counted in INSPIRE as of 03 Jun 2021

%\cite{Braaten:1993rw}
\bibitem{Braaten:1993rw}
E.~Braaten and T.~C.~Yuan,
%``Gluon fragmentation into heavy quarkonium,''
Phys. Rev. Lett. \textbf{71}, 1673-1676 (1993)
doi:10.1103/PhysRevLett.71.1673
[arXiv:hep-ph/9303205 [hep-ph]].
%321 citations counted in INSPIRE as of 03 Jun 2021


%\cite{Braaten:1995cj}
\bibitem{Braaten:1995cj}
E.~Braaten and T.~C.~Yuan,
%``Gluon fragmentation into spin triplet S wave quarkonium,''
Phys. Rev. D \textbf{52}, 6627-6629 (1995)
doi:10.1103/PhysRevD.52.6627
[arXiv:hep-ph/9507398 [hep-ph]].
%89 citations counted in INSPIRE as of 03 Jun 2021

%\cite{Cho:1994gb}
\bibitem{Cho:1994gb}
P.~L.~Cho, M.~B.~Wise and S.~P.~Trivedi,
%``Gluon fragmentation into polarized charmonium,''
Phys. Rev. D \textbf{51}, R2039-R2043 (1995)
doi:10.1103/PhysRevD.51.R2039
[arXiv:hep-ph/9408352 [hep-ph]].
%83 citations counted in INSPIRE as of 03 Jun 2021

%\cite{Zhang:2017xoj}
\bibitem{Zhang:2017xoj}
P.~Zhang, Y.~Q.~Ma, Q.~Chen and K.~T.~Chao,
%``Analytical calculation for the gluon fragmentation into spin-triplet S-wave quarkonium,''
Phys. Rev. D \textbf{96}, no.9, 094016 (2017)
doi:10.1103/PhysRevD.96.094016
[arXiv:1708.01129 [hep-ph]].
%7 citations counted in INSPIRE as of 03 Jun 2021

%\cite{Bodwin:2003wh}
\bibitem{Bodwin:2003wh}
G.~T.~Bodwin and J.~Lee,
%``Relativistic corrections to gluon fragmentation into spin triplet S wave quarkonium,''
Phys. Rev. D \textbf{69}, 054003 (2004)
doi:10.1103/PhysRevD.69.054003
[arXiv:hep-ph/0308016 [hep-ph]].
%37 citations counted in INSPIRE as of 03 Jun 2021

%\cite{Bodwin:2012xc}
\bibitem{Bodwin:2012xc}
G.~T.~Bodwin, U.~R.~Kim and J.~Lee,
%``Higher-order relativistic corrections to gluon fragmentation into spin-triplet S-wave quarkonium,''
JHEP \textbf{11}, 020 (2012)
doi:10.1007/JHEP11(2012)020
[arXiv:1208.5301 [hep-ph]].
%32 citations counted in INSPIRE as of 03 Jun 2021

%\cite{Gao:2016ihc}
\bibitem{Gao:2016ihc}
X.~Gao, Y.~Jia, L.~Li and X.~Xiong,
%``Relativistic correction to gluon fragmentation function into pseudoscalar quarkonium,''
Chin. Phys. C \textbf{41}, no.2, 023103 (2017)
doi:10.1088/1674-1137/41/2/023103
[arXiv:1606.07455 [hep-ph]].
%6 citations counted in INSPIRE as of 03 Jun 2021

%\cite{Beneke:1995yb}
\bibitem{Beneke:1995yb}
M.~Beneke and I.~Z.~Rothstein,
%``Psi-prime polarization as a test of color octet quarkonium production,''
Phys. Lett. B \textbf{372}, 157-164 (1996)
[erratum: Phys. Lett. B \textbf{389}, 769 (1996)]
doi:10.1016/0370-2693(96)00030-5
[arXiv:hep-ph/9509375 [hep-ph]].
%117 citations counted in INSPIRE as of 03 Jun 2021

%\cite{Braaten:2000pc}
\bibitem{Braaten:2000pc}
E.~Braaten and J.~Lee,
%``Next-to-leading order calculation of the color octet 3S(1) gluon fragmentation function for heavy quarkonium,''
Nucl. Phys. B \textbf{586}, 427-439 (2000)
doi:10.1016/S0550-3213(00)00396-5
[arXiv:hep-ph/0004228 [hep-ph]].
%52 citations counted in INSPIRE as of 03 Jun 2021

%\cite{Artoisenet:2014lpa}
\bibitem{Artoisenet:2014lpa}
P.~Artoisenet and E.~Braaten,
%``Gluon fragmentation into quarkonium at next-to-leading order,''
JHEP \textbf{04}, 121 (2015)
doi:10.1007/JHEP04(2015)121
[arXiv:1412.3834 [hep-ph]].
%17 citations counted in INSPIRE as of 03 Jun 2021

%\cite{Artoisenet:2018dbs}
\bibitem{Artoisenet:2018dbs}
P.~Artoisenet and E.~Braaten,
%``Gluon fragmentation into quarkonium at next-to-leading order using FKS subtraction,''
JHEP \textbf{01}, 227 (2019)
doi:10.1007/JHEP01(2019)227
[arXiv:1810.02448 [hep-ph]].
%12 citations counted in INSPIRE as of 03 Jun 2021

%\cite{Feng:2018ulg}
\bibitem{Feng:2018ulg}
F.~Feng and Y.~Jia,
%``Next-to-leading-order QCD corrections to gluon fragmentation into ${}^1S_0^{(1,8)}$ quarkonia,''
[arXiv:1810.04138 [hep-ph]].
%9 citations counted in INSPIRE as of 03 Jun 2021

%\cite{Zhang:2018mlo}
\bibitem{Zhang:2018mlo}
P.~Zhang, C.~Y.~Wang, X.~Liu, Y.~Q.~Ma, C.~Meng and K.~T.~Chao,
%``Semi-analytical calculation of gluon fragmentation into$^{1}$S$_{0}^{[1,8]}$ quarkonia at next-to-leading order,''
JHEP \textbf{04}, 116 (2019)
doi:10.1007/JHEP04(2019)116
[arXiv:1810.07656 [hep-ph]].
%13 citations counted in INSPIRE as of 03 Jun 2021

%\cite{Braaten:1994kd}
\bibitem{Braaten:1994kd}
E.~Braaten and T.~C.~Yuan,
%``Gluon fragmentation into P wave heavy quarkonium,''
Phys. Rev. D \textbf{50}, 3176-3180 (1994)
doi:10.1103/PhysRevD.50.3176
[arXiv:hep-ph/9403401 [hep-ph]].
%115 citations counted in INSPIRE as of 03 Jun 2021

%\cite{Ma:1995ci}
\bibitem{Ma:1995ci}
J.~P.~Ma,
%``Gluon fragmentation into P wave triplet quarkonium,''
Nucl. Phys. B \textbf{447}, 405-424 (1995)
doi:10.1016/0550-3213(95)00270-3
[arXiv:hep-ph/9503346 [hep-ph]].
%42 citations counted in INSPIRE as of 03 Jun 2021

%\cite{Hao:2009fa}
\bibitem{Hao:2009fa}
G.~Hao, Y.~Zuo and C.~F.~Qiao,
%``The Fragmentation Function of Gluon Splitting into P-wave Spin-singlet Heavy Quarkonium,''
[arXiv:0911.5539 [hep-ph]].
%10 citations counted in INSPIRE as of 03 Jun 2021

%\cite{Feng:2017cjk}
\bibitem{Feng:2017cjk}
F.~Feng, S.~Ishaq, Y.~Jia and J.~Y.~Zhang,
%``Fragmentation function of gluon into spin-singlet $P$-wave quarkonium,''
Phys. Rev. D \textbf{102}, no.1, 014038 (2020)
doi:10.1103/PhysRevD.102.014038
[arXiv:1712.09986 [hep-ph]].
%6 citations counted in INSPIRE as of 03 Jun 2021

%\cite{Zhang:2020atv}
\bibitem{Zhang:2020atv}
P.~Zhang, C.~Meng, Y.~Q.~Ma and K.~T.~Chao,
%``Gluon fragmentation into ${^{3}\hspace{-0.6mm}P_{J}^{[1,8]}}$ quark pair and test of NRQCD factorization at two-loop level,''
[arXiv:2011.04905 [hep-ph]].
%1 citations counted in INSPIRE as of 03 Jun 2021

%\cite{Collins:1981uw}
\bibitem{Collins:1981uw}
J.~C.~Collins and D.~E.~Soper,
%``Parton Distribution and Decay Functions,''
Nucl. Phys. B \textbf{194}, 445-492 (1982)
doi:10.1016/0550-3213(82)90021-9
%1079 citations counted in INSPIRE as of 03 Jun 2021

%\cite{Petrelli:1997ge}
\bibitem{Petrelli:1997ge}
A.~Petrelli, M.~Cacciari, M.~Greco, F.~Maltoni and M.~L.~Mangano,
%``NLO production and decay of quarkonium,''
Nucl. Phys. B \textbf{514}, 245-309 (1998)
doi:10.1016/S0550-3213(97)00801-8
[arXiv:hep-ph/9707223 [hep-ph]].
%278 citations counted in INSPIRE as of 03 Jun 2021

%\cite{Nogueira:1991ex}
\bibitem{Nogueira:1991ex}
P.~Nogueira,
%``Automatic Feynman graph generation,''
J. Comput. Phys. \textbf{105}, 279-289 (1993)
doi:10.1006/jcph.1993.1074
%1000 citations counted in INSPIRE as of 03 Jun 2021

%\cite{Hahn:2000kx}
\bibitem{Hahn:2000kx}
T.~Hahn,
%``Generating Feynman diagrams and amplitudes with FeynArts 3,''
Comput. Phys. Commun. \textbf{140}, 418-431 (2001)
doi:10.1016/S0010-4655(01)00290-9
[arXiv:hep-ph/0012260 [hep-ph]].
%1722 citations counted in INSPIRE as of 03 Jun 2021

%\cite{Mertig:1990an}
\bibitem{Mertig:1990an}
R.~Mertig, M.~Bohm and A.~Denner,
%``FEYN CALC: Computer algebraic calculation of Feynman amplitudes,''
Comput. Phys. Commun. \textbf{64}, 345-359 (1991)
doi:10.1016/0010-4655(91)90130-D
%1016 citations counted in INSPIRE as of 03 Jun 2021

%\cite{Feng:2012tk}
\bibitem{Feng:2012tk}
F.~Feng and R.~Mertig,
%``FormLink/FeynCalcFormLink : Embedding FORM in Mathematica and FeynCalc,''
[arXiv:1212.3522 [hep-ph]].
%42 citations counted in INSPIRE as of 03 Jun 2021

%\cite{Feng:2012iq}
\bibitem{Feng:2012iq}
F.~Feng,
%``$\tt{Apart}$: A Generalized Mathematica Apart Function,''
Comput. Phys. Commun. \textbf{183}, 2158-2164 (2012)
doi:10.1016/j.cpc.2012.03.025
[arXiv:1204.2314 [hep-ph]].
%60 citations counted in INSPIRE as of 03 Jun 2021

%\cite{Binoth:2000ps}
\bibitem{Binoth:2000ps}
T.~Binoth and G.~Heinrich,
%``An automatized algorithm to compute infrared divergent multiloop integrals,''
Nucl. Phys. B \textbf{585}, 741-759 (2000)
doi:10.1016/S0550-3213(00)00429-6
[arXiv:hep-ph/0004013 [hep-ph]].
%452 citations counted in INSPIRE as of 03 Jun 2021

%\cite{Binoth:2003ak}
\bibitem{Binoth:2003ak}
T.~Binoth and G.~Heinrich,
%``Numerical evaluation of multiloop integrals by sector decomposition,''
Nucl. Phys. B \textbf{680}, 375-388 (2004)
doi:10.1016/j.nuclphysb.2003.12.023
[arXiv:hep-ph/0305234 [hep-ph]].
%188 citations counted in INSPIRE as of 03 Jun 2021


%\cite{Chetyrkin:2000yt}
\bibitem{Chetyrkin:2000yt}
K.~G.~Chetyrkin, J.~H.~Kuhn and M.~Steinhauser,
%``RunDec: A Mathematica package for running and decoupling of the strong coupling and quark masses,''
Comput. Phys. Commun. \textbf{133}, 43-65 (2000)
%doi:10.1016/S0010-4655(00)00155-7
[arXiv:hep-ph/0004189 [hep-ph]].
%494 citations counted in INSPIRE as of 03 Jun 2021

%\cite{Zheng:2021ylc}
\bibitem{Zheng:2021ylc}
X.~C.~Zheng, X.~G.~Wu and X.~D.~Huang,
%``NLO fragmentation functions for a quark into a spin-singlet quarkonium: Same flavor case,''
[arXiv:2105.14580 [hep-ph]].
%0 citations counted in INSPIRE as of 03 Jun 2021

\end{thebibliography}
\end{document}